\documentclass[aps,amsmath,amssymb,prl,twocolumn,preprintnumbers,superscriptaddress,nobalancelastpage]{revtex4}
\usepackage{graphicx}

\begin{document}
\title{
Magnetic-field-induced spin excitations and renormalized  spin gap of the underdoped 
superconductor La$_{1.895}$Sr$_{0.105}$CuO$_{4}$ 
}

\author{
J.\ Chang
       }
\affiliation{
Laboratory for Neutron Scattering, ETH Zurich and PSI Villigen, CH-5232 Villigen PSI, Switzerland
            }
\author{
A.P.\ Schnyder
       }
\affiliation{
Condensed Matter Theory Group, Paul Scherrer Institute, CH-5232 Villigen PSI, Switzerland
            }
\author{
R.\ Gilardi
       }
\affiliation{
Laboratory for Neutron Scattering, ETH Zurich and PSI Villigen, CH-5232 Villigen PSI, Switzerland
            }
\author{
H.M.\ R\o nnow
       }
\affiliation{
Laboratory for Neutron Scattering, ETH Zurich and PSI Villigen, CH-5232 Villigen PSI, Switzerland
            }
\author{
S.\ Pailhes
       }
\affiliation{
Laboratory for Neutron Scattering, ETH Zurich and PSI Villigen, CH-5232 Villigen PSI, Switzerland
            }
\author{
N.B.\ Christensen
       }
\affiliation{
Laboratory for Neutron Scattering, ETH Zurich and PSI Villigen, CH-5232 Villigen PSI, Switzerland
            }
\affiliation{
Ris\o\ National Laboratory, DK-4000 Roskilde, Denmark
            }
\author{
Ch.\  Niedermayer
       }
\affiliation{
Laboratory for Neutron Scattering, ETH Zurich and PSI Villigen, CH-5232 Villigen PSI, Switzerland
            }
\author{
D.F.\ McMorrow
       }
\affiliation{
London Centre for Nanotechnology and Department of Physics and Astronomy,  University College London, London, UK
            }
\affiliation{ISIS Facility, Rutherford Appleton Laboratory, Chilton, Didcot OX11 0QX, UK}
\author{
A.\ Hiess
       }
\affiliation{
Institut Laue-Langevin, BP 156, F-38042 Grenoble, France
            }
\author{
A.\ Stunault
       }
\affiliation{
Institut Laue-Langevin, BP 156, F-38042 Grenoble, France
            }
\author{
M.\ Enderle
       }
\affiliation{
Institut Laue-Langevin, BP 156, F-38042 Grenoble, France
            }
\author{
B.\ Lake
       }
\affiliation{
BENSC Hahn-Meitner-Institut, 14109 Berlin Wannsee, Germany
            }
\author{
O.\ Sobolev
       }
\affiliation{
BENSC Hahn-Meitner-Institut, 14109 Berlin Wannsee, Germany
            }
\author{
N.\ Momono
       }
\affiliation{
Department of Physics, Hokkaido University - Sapporo 060-0810, Japan
            }
\author{
M.\ Oda
       }
\affiliation{
Department of Physics, Hokkaido University - Sapporo 060-0810, Japan
            }
\author{
M.\ Ido
       }
\affiliation{
Department of Physics, Hokkaido University - Sapporo 060-0810, Japan
            }
\author{
C.\ Mudry
       }
\affiliation{
Condensed Matter Theory Group, Paul Scherrer Institute, CH-5232 Villigen PSI, Switzerland
            }
\author{
J.\ Mesot
       }
\email{joel.mesot@psi.ch}
\affiliation{
Laboratory for Neutron Scattering, ETH Zurich and PSI Villigen, CH-5232 Villigen PSI, Switzerland
            }

\begin{abstract}
High-resolution neutron inelastic scattering experiments 
in applied magnetic fields have been performed on 
La$_{1.895}$Sr$_{0.105}$CuO$_{4}$ (LSCO). In zero field, 
the temperature dependence of the low-energy peak intensity at the
incommensurate momentum-transfer
$\mathbf{Q}^{\ }_{\mathrm{IC}}=(0.5,0.5\pm\delta,0),(0.5\pm\delta,0.5,0)$
exhibits an anomaly at the superconducting $T^{\ }_{c}$ which
broadens and shifts to lower temperature upon the application of a
magnetic field along the c-axis. A field-induced
enhancement of the spectral weight is observed, but only at finite
energy transfers and in an intermediate temperature range. 
These observations establish the opening of a strongly
downward renormalized spin gap in the underdoped regime of LSCO.
This behavior contrasts with the observed doping dependence of most
electronic energy features. 
\end{abstract}
\date{\today}
\maketitle

In studies of high-temperature superconductors (HTSC), one central challenge is 
to explain the evolution from an antiferromagnetic (AF) Mott
insulator to a metallic superconductor upon doping.
For example, the intimate interplay between magnetism and superconductivity
has been revealed in momentum resolved inelastic neutron scattering (INS) experiments
on La$_{2-x}$Sr$_{x}$CuO$_{4}$ (LSCO)%
~\cite{masonprl92,yamada95,Gilardi04epl,bella99,waki}
and YBa$_2$Cu$_3$O$_{6+x}$ (YBCO)%
~\cite{bourgesbook,daisc99}
that showed
the opening of a
spin gap (SG) $\Delta^{\ }_{sg}$
at the superconducting (SC) critical temperature ($T^{\ }_{c}$).
In contrast to the single-particle SC gap probed
by, e.g., angle-resolved photoemission spectroscopy (ARPES)%
~\cite{campuzano04,damascelli03}, the SG scales with $T^{\ }_{c}$ in
the overdoped to slightly underdoped regime. The situation is more
complicated in more  underdoped samples. For instance, in YBCO the
ratio $\Delta^{\ }_{sg}/(k^{\ }_{B}T^{\ }_{c})$ decreases with strong
underdoping~\cite{bourgesgap,daiprb01}, while for underdoped LSCO
($x<$0.12) the changes are even more dramatic, since no direct
neutron scattering evidence for a SG has been reported so far
\cite{Lee}. This latter fact would seem to indicate that the
underdoped regime of LSCO cannot be described in terms of a
homogeneous electronic liquid and more exotic scenarios have been
suggested, such as the formation of dynamical
stripes~\cite{kivelsonRevmod} or presence of a $d$-density wave
(DDW) order \cite{ddw}. Further insight into the interplay between
magnetism and superconductivity in LSCO 
has been revealed in a number of studies of the effects of 
an applied magnetic field, which at optimal doping enhances the 
low-energy magnetic fluctuations and at underdoping induces 
static antiferromagnetism \cite{lake01,lakenature02,demler}.

In this paper, based on high-resolution
INS data as a function of energy $\hbar\omega$, 
magnetic field $H$ and temperature $T$, we identify similarities and
differences between the magnetic response in the underdoped and
optimal doped regimes of LSCO. Our main result is the observation of a dramatic
field induced enhancement of the response at low energies and the identification,
below $T^{\ }_{c}$, of a SG in the  underdoped regime ($x=0.105$),
whose characteristic energy is strongly renormalized relative to
that measured at optimal doping. We also show that this behavior
is not specific to LSCO but corresponds to what is observed in
YBCO. Unlike at optimal doping, the opening of the $x=0.105$ SG is
incomplete all the way down to the lowest achievable temperature.

\begin{figure}
\includegraphics[width=0.4\textwidth]{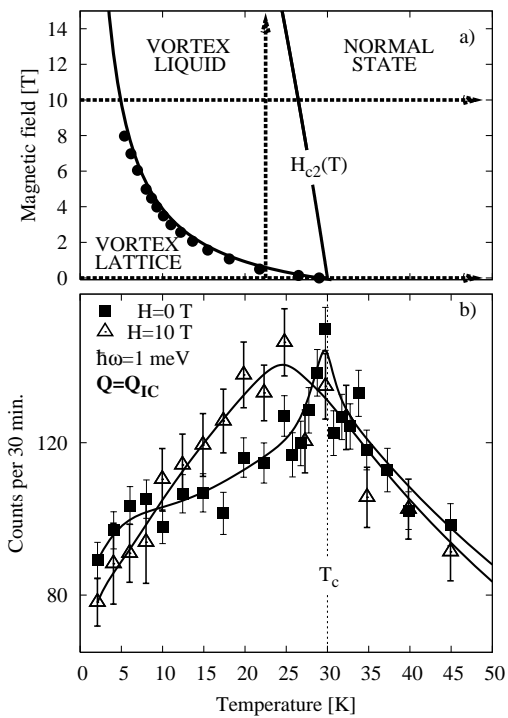}
\caption{
(a) Phase diagram of underdoped LSCO ($x=0.105$)
as a function of a magnetic field $H$
 and temperature $T$~\cite{Gilardi05}.
Full circles are the measured values
for the melting line of the vortex lattice~\cite{Gilardi05}.
The straight line is the position of $T^{\ }_{c}$ as a function of $H$.
The dashed arrows indicate the $T$ and $H$ scans
that are described in this paper.
(b) Two $T$ scans at $\mathbf{Q}^{\ }_{\mathrm{IC}}$
and $\hbar\omega=1\, \mathrm{meV}$ transfer. Open triangles
(filled squares) denote the measured intensity at $H=10\, \mathrm{T}$
($H=0\, \mathrm{T}$),
solid lines are guides to the eye.}
\label{fig:fig1}
\end{figure}

Our experiments have been performed on a large cylindrical single crystal of
La$_{2-x}$Sr$_{x}$CuO$_4$ ($m\sim3.86\, \mathrm{g}$, $x=0.105$) grown by the
traveling solvent floating zone (TSFZ) method~\cite{tsfz}. The SC volume fraction
 was estimated
from measurements of the specific heat $C$. At $T^{\ }_{c}=30\, \mathrm{K}$,
we observed a jump of
$\Delta C/T^{\ }_{c}=3\, \mathrm{mJ/(K}^2\mathrm{mol})$.
Below $T^{\ }_{c}$ the electronic contribution
was estimated to be
$C_{el}^s/T=\gamma_{el}^s=0.5\, \mathrm{mJ/(K}^2\mathrm{mol})$
while in the normal state we found
$C_{el}^n/T=\gamma_{el}^n=5.5\, \mathrm{mJ/(K}^2\mathrm{mol})$.
This indicates that the SC volume fraction
$1- \gamma_{el}^s/\gamma_{el}^n$ is larger than 90$\%$.
Together with a sharp SC transition
($\Delta T^{\ }_{c}=1.5\, \mathrm{K}$),
this demonstrates the high quality of the crystal.
The structural transition from the high-temperature tetragonal to the
low-temperature orthorhombic phase occurs at $290\, \mathrm{K}$, which
confirms $x=0.105\pm0.005$.
The magnetic response is peaked in the reciprocal space
 at the incommensurate wave vector
$\mathbf{Q}^{\ }_{\mathrm{IC}}=(0.5\pm\delta,0.5,0),(0.5,0.5\pm\delta,0)$
(in tetragonal units of $2\pi/a=1.65\ \AA^{-1}$ used throughout this paper).
The  value $\delta=0.097\pm0.005$ is in line with
a Sr concentration of $x=0.105\pm0.005$ \cite{yamadadelta}.
The sample has been characterized by magnetization and ac-susceptibility
measurements in magnetic fields $H$ up to $8\, \mathrm{T}$
 along the crystallographic $c$ axis~\cite{Gilardi05},
as well as small-angle neutron scattering (SANS) and $\mu$SR~\cite{Divakar04}.
All measurements indicate that in the  underdoped regime of LSCO,
the vortex liquid phase dominates the magnetic phase diagram,
while a quasi-long-range ordered vortex lattice
can only be observed at low fields/temperatures~\cite{Divakar04}.
This contrasts with the well-ordered vortex lattice observed at
all fields in the low-temperature regime of optimally doped LSCO%
~\cite{Gilardi02}. The melting line of the vortex lattice is shown in
Fig.~\ref{fig:fig1}a.

%%%%%%%%%%FIG 1%%%%%%%%%%%%%%%
%%%%%%%%%END FIG 1%%%%%%%%%%%%%%%%

%%%%%%%%%%FIG 2%%%%%%%%%%%%%%%
\begin{figure}
\includegraphics[width=0.37\textwidth]{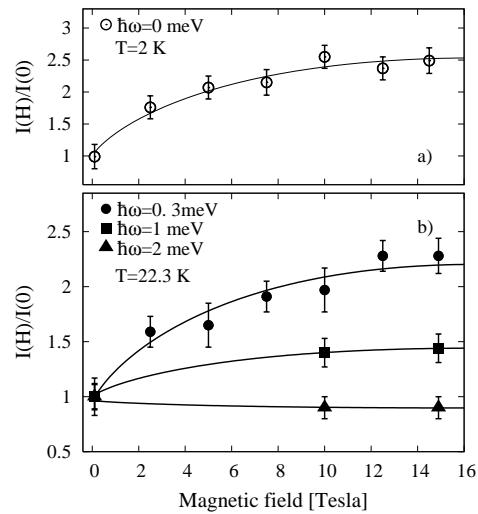}
\caption{ Magnetic field dependence of the response at  $\mathbf{Q}^{\ }_{\mathrm{IC}}$ for LSCO ($x=0.105$).
(a) Elastic response with $T=2\  \mathrm{K}$. The solid line is a theoretical prediction \cite{demler}.
(b) Inelastic response with $T=22.3\ \mathrm{K}$, 
    solid lines are guides to the eye.}
\label{fig:fig2}
\end{figure}
%%%%%%%%%END FIG 2%%%%%%%%%%%%%%%%

INS measurements of the spin excitations
were performed on the cold neutron spectrometer IN14 at the
Institut Laue Langevin, Grenoble, France.
We used a vertically curved graphite monochromator and a
horizontally curved graphite analyzer with fixed final energy $E_f$ in the range $3.5-5\ \mathrm{meV}$ that gave an energy resolution of about
$60-150\ \mu\mathrm{eV}$.
A cooled Beryllium-filter  removed higher-order
contamination from the scattered beam. %$\mathbf{k}_{f}$.
The sample was cut into two pieces which were coaligned within
less than a degree and mounted in a 15 $\mathrm{T}$ vertical cryomagnet such 
that momentum transfers $\textbf{Q}=(Q_h,Q_k,0)$
were accessible.
All measurements were performed after field cooling from above $T^{\ }_{c}$.
With a similar setup the field dependence of the elastic signal at
the incommensurate wave vectors was measured on the FLEX spectrometer at Hahn-Meitner Institute.
  
In Fig.\
\ref{fig:fig1}b, we  show  the effect of
$H$ on the $T$ dependence of the intensity at
$\mathbf{Q}^{\ }_{\mathrm{IC}}$
with an energy transfer of $\hbar\omega=1\ \mathrm{meV}$. These data
demonstrate that the  intensity is
enhanced by application of $H=10\, \mathrm{T}$ in an intermediated temperature range
$10\, \mathrm{K}<T<T^{\ }_{c}$, corresponding roughly to the liquid region of the vortex 
phase diagram, see Fig.~\ref{fig:fig1}a. A similar behavior was previously reported for LSCO 
samples close to optimal doping \cite{Gilardi04epl,lake01}. This was interpreted in terms of 
induced excitations inside the vortex core \cite{lake01}.
In effect, $H$ shifts and broadens the cusp-like anomaly at
$T^{\ }_{c}$ to a lower $T^{\ }_{c}(H)$. The shift corresponds qualitatively
to the shift in $T^{\ }_{c}$ as seen in Fig.~\ref{fig:fig1}a.
We note that 
the lower limit on the field effect might be related to the onset 
at  $T^{\ }_{f}=10\, \mathrm{K}$ of  
Cu moment freezing observed by  $\mu$SR on the same sample~\cite{chang}.

%%%%%%%%%%%%%%%%%%FIG 3%%%%%%%%%%%%%%%%%%%
\begin{figure}
\hspace{-1mm}
\includegraphics[width=0.49\textwidth]{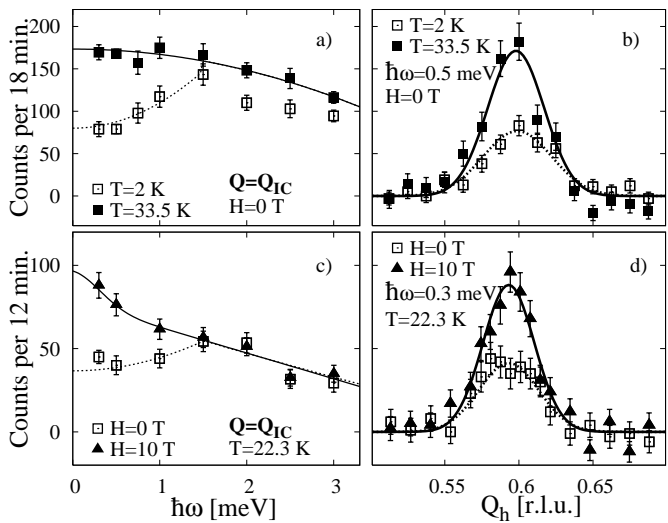}
\caption{Energy and  $\mathbf{Q}$ dependence of magnetic scattering from LSCO ($x=0.105$). 
Energy scans are taken at  $\mathbf{Q}^{\ }_{\mathrm{IC}}$:
(a) Comparison of the response in the SC phase with that of the normal state in zero field.
(c) Effect of a $10\, \mathrm{T}$ field at $T=22.3\ \mathrm{K}$.
Q scans through $\mathbf{Q}^{\ }_{\mathrm{IC}}$:
(b) Comparison of the response in the SC phase with that of the normal state using $\hbar\omega=$0.5 meV.
(d) The effect of a $10\, \mathrm{T}$ field at $22.3\ \mathrm{K}$,  $\hbar\omega=$0.3 meV, $E^{\ }_f=$3.5 meV. 
In this configuration the resolution (HWHM) was 0.03 meV. The data have been background 
subtracted and lines are guide to the eye.}
\label{fig:fig3}
\end{figure}
%%%%%%%%%%%% END Fig 3%%%%%%%%%%%%%%%%%%%%%%

Next, we plot in Fig.~\ref{fig:fig2} the $H$ dependence of the
normalized intensity  at
$\mathbf{Q}^{\ }_{\mathrm{IC}}$. 
The \emph{elastic} magnetic signal at $T=2$ K 
exhibits a large enhancement in agreement with previous reports on samples of
similar doping levels~\cite{lakenature02}, and is well fitted by the equation  
$I\sim (H/H^{\ }_{c2})\ln(H^{\ }_{c2}/H)$ predicted by Demler \textit{et al.}%
~\cite{demler}. 
Remarkably, the \emph{inelastic} response at $T$=23 K exhibits a field 
dependence at low energy transfers similar to the \emph{elastic} signal.
The dependence on $\hbar\omega$ and $\mathbf{Q}$ of the intensity, 
taken at different temperatures and fields,
is shown in Fig.~\ref{fig:fig3}.
In Fig.~\ref{fig:fig3}a the  intensity at
$\mathbf{Q}^{\ }_{\mathrm{IC}}$,
and at $H=0\ \mathrm{T}$ shows a smooth dependence on $\hbar\omega$
above  $T^{\ }_{c}$.
At $T=2\ \mathrm{K}$,
an anomaly is exhibited at $\hbar\omega=1.5\ \mathrm{meV}$
below which intensity is reduced relative to that in the normal state.
Figure~\ref{fig:fig3}b
shows the effect of cooling  through $T^{\ }_{c}$ in the $\mathbf{Q}$ scan
through $\mathbf{Q}^{\ }_{IC}$
 at $H=0\ \mathrm{T}$
and with $\hbar\omega=0.5\ \mathrm{meV}$.
Cooling  from $33.5\ \mathrm{K}$ to $2\ \mathrm{K}$
results in a two-fold reduction of the intensity at $\mathbf{Q}^{\ }_{\mathrm{IC}}$. 
Figures~\ref{fig:fig3}c and \ref{fig:fig3}d show
the change in the dependence on $\hbar\omega$ and $\mathbf{Q}$ of the intensity
 at fixed $T=22.3\ \mathrm{K}$ upon applying $H=10\, \mathrm{T}$, respectively.
The effect of the applied field,  Fig.~\ref{fig:fig3}c, is to
remove the zero-field anomaly at $\hbar\omega=1.5\ \mathrm{meV}$,
and is similar to the effect of raising the temperature above
$T^{\ }_{c}$, Fig.~\ref{fig:fig3}a. Note that here the
magnetic-field-induced redistribution of spectral weight is
limited to frequencies smaller than $1.5\ \mathrm{meV}$.

To better appreciate the features of the excitation spectrum
described above, a comparison with the SG signatures observed in
optimally doped LSCO is relevant. (i) Although the details vary
from experiment to experiment
\cite{lake01,tranquadaprb04,Gilardi04epl}, magnetic field-induced
excitations for $\hbar\omega<\Delta^{\ }_{sg}$ and $T<T^{\ }_{c}$ remain a
common observation. (ii) For $\hbar\omega<\Delta^{\ }_{sg}$, beside the
suppression of intensity at low temperatures, an anomaly was
reported at $T^{\ }_{c}$ that marks the closing of the spin gap as the
temperature crosses $T^{\ }_{c}$ \cite{mason96}. (iii) A strong
suppression of the intensity below
 $\hbar\omega\approx\Delta^{\ }_{sg}$
is observed in the SC state but not in the normal state%
~\cite{christensen}.
In our underdoped LSCO sample, we infer from
Figs.~\ref{fig:fig1}b and \ref{fig:fig3}
similar spin gap features as in optimally doped LSCO.
(i) An applied magnetic field acts to redistribute the spectral weight for
$10\, \mathrm{K}<T<T^{\ }_{c}$ at $\hbar\omega<1.5\ \mathrm{meV}$ 
(Fig.~\ref{fig:fig1}b and~\ref{fig:fig3}c).
(ii) There is the cusp-like anomaly at  $T^{\ }_{c}$ scanning the temperature
for fixed energy below  $1.5\ \mathrm{meV}$.
(iii) Energy scans reveal an anomaly at $\hbar\omega=1.5\ \mathrm{meV}$ 
which disappears in the normal state.
The field effect (i) alone provides compelling evidence for a spin
gap at 1.5 meV. This is further supported by the cusp at $T^{\ }_{c}$
(ii), and consistent with  the energy dependence (iii).
We also find that the intensity at $\mathbf{Q}^{\ }_{\mathrm{IC}}$
displays
the same 
functional dependence as that reported near optimal 
doping \cite{aeppli97} as a function of $T>T^{\ }_{c}$ or $\hbar\omega>1.5\ \mathrm{meV}$.
%%%%%%%%%%%%%%FIG 4%%%%%%%%%%%%%%
\begin{figure}
\includegraphics[width=0.499\textwidth]{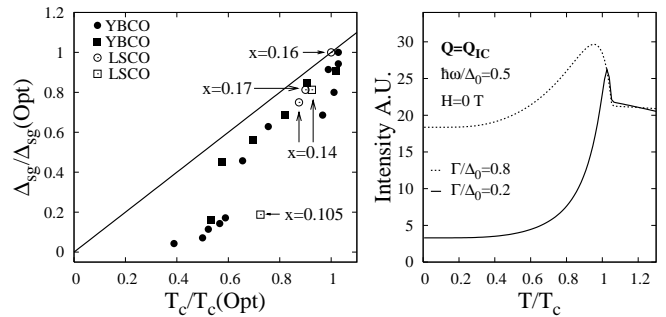}
\caption{(a) Normalized $\Delta^{\ }_{sg}$ versus normalized $T^{\ }_{c}$
for LSCO (open symbols) and YBCO (filled circles \cite{bourgesgap}
and filled squares \cite{daiprb01}). The open circles are taken
from previously published data, $x=\text{0.14 \cite{mason96}}$,
$x=\text{0.16 \cite{bella99}}$ and $x=\text{0.17
\cite{Gilardi04epl}}$ while open squares are from this work
($x=\text{0.105}$) and  Ref. \cite{chang} $(x=\text{0.14})$. (b)
Intensity at $\mathbf{Q}^{\ }_{\mathrm{IC}}$, at the energy transfer
$\hbar\omega=0.5\Delta^{\ }_{0}$, and at $H=0\ \mathrm{T}$ versus $T$
calculated within a phenomenological Fermi-liquid (PFL)
approach. The solid (dashed) line uses the
inverse lifetime $\Gamma=0.2\Delta^{\ }_{0}$
($\Gamma=0.8\Delta^{\ }_{0}$). }\label{fig:fig4}  \label{fig:fig5}
\end{figure}

We plot in Fig.~\ref{fig:fig4}a the dependence of the SG on $T^{\ }_{c}$
for LSCO and YBCO samples. Vertical and horizontal axes have been
normalized by the values of the SG and SC critical temperature at
optimal doping, respectively. The ratio 
$\Delta^{\ }_{sg}/(k^{\ }_{B}T^{\ }_{c})$
exhibits a strong downward multiplicative renormalization in the
underdoped regime. This strong renormalization was also observed
in YBCO\ \cite{bourgesgap,daiprb01}, in which context it was
attributed to inhomogeneities induced by oxygen doping\
\cite{daiprb01}. We must, however, rule out this explanation for
our $x=0.105$ LSCO sample in view of the sharpness of the SC
transition ($\Delta T^{\ }_{c}=1.5\ \mathrm{K}$). Instead, we believe
that the proper interpretation of Fig.~\ref{fig:fig4} is that of a
universal doping dependence of
$\Delta^{\ }_{sg}/\Delta^{\ }_{sg}(\mathrm{Opt})$ with a linear regime that
defines the approach to optimal doping and an underdoped regime
whose defining property is a doping dependent
$\Delta^{\ }_{sg}/(k^{\ }_{B}T^{\ }_{c})$.

It is remarkable that $\Delta^{\ }_{sg}/(k^{\ }_{B}T^{\ }_{c})$ 
decreases in the underdoped regime, 
since this is opposite to the doping dependence
of most energy features (maximum SC gap or pseudogap) which are
found, from ARPES measurements, to increase with underdoping
\cite{campuzano04,damascelli03}.
In a phenomenological Fermi-liquid (PFL) picture, 
which is relatively successful at optimal
doping, the SG is directly related to the SC gap.
Although its maximum increases with underdoping, the slope of the
SC $d$-wave gap has been measured to soften at the nodes with
underdoping \cite{softening d-wave slope}. This means that the
ratio $\Delta^{\ }_{sg}/(k^{\ }_{B}T^{\ }_{c})$ will decrease with underdoping%
~\cite{schnyder}.
In this picture, the anomaly at $T^{\ }_{c}$ in the temperature
dependence displayed in Fig.~\ref{fig:fig1}b is caused by the
closing of the SG upon entering the normal state. In
Fig.~\ref{fig:fig4}b we show the calculated dependence on
temperature of the zero field neutron intensity at
$\mathbf{Q}^{\ }_{\mathrm{IC}}$ and energy transfer 
$\hbar\omega=0.5\Delta^{\ }_{0}$
using the PFL parameters chosen to reproduce 
the $T$ dependence at optimal doping~\cite{footnote}.
A cusp-like anomaly
at $T^{\ }_{c}$ separates the normal regime from a $d$-wave SC regime
with a decreasing residual intensity due to a phenomenological
inverse life-time of $\Gamma=0.2\Delta^{\ }_{0}$. In this spirit, to
reproduce the measured residual spectral weight at low $T$ in
$x=0.105$ LSCO would require $\Gamma\approx\Delta^{\ }_{sg}$ resulting
in strong broadening of the peak at $T^{\ }_{c}$ as shown in
Fig.~\ref{fig:fig4}b. Thus, the naive PFL approach is unable to
reconcile a sharp anomaly at $T^{\ }_{c}$ ($\Delta^{\ }_{sg}$) for small
enough $\hbar\omega$ ($T$) and the large, compared to optimal
doping, residual intensity in the low-$T$ (low-$\hbar\omega$) tail
of Fig.~\ref{fig:fig1}b (Fig.~\ref{fig:fig3}a).

One way to explain the large residual intensity at low
temperatures and energies in strongly underdoped HTSC could be in
terms of a two-component model. The first component would
correspond to the gapped response seen around optimal doping. The
second (quasi-elastic) component could be related to the slowing
down of the magnetic signal observed by $\mu$SR~\cite{muSR sees
slowing down}, and the central mode recently reported in YBCO
($T^{\ }_{c}=18\ \mathrm{K}$)~\cite{stock}.
The fact that both components 
develop well-defined peaks at $\mathbf{Q}^{\ }_{\mathrm{IC}}$ suggests
an inter-twinning of the two components on a nanometer length
scale. This could possibly be explained within a picture of static
and fluctuating stripes, although it remains a challenge to
quantify this picture to describe the detailed field and
temperature effects of the magnetic response reported here.

In summary, we have reported a dramatic field-induced enhancement
of the INS response for low energy transfers in a restricted
temperature range. The field dependence suggests that the approach
developed \cite{demler} to explain the previously observed 
field-induced static magnetic order \cite{lake01}
(see also Ref.\ \onlinecite{lakenature02}) may also be
applicable to the dynamics. The field-induced excitation may be
more difficult to capture in a PFL approach. Furthermore, we have
identified a SG, $\Delta^{\ }_{sg}=1.5$ meV in underdoped LSCO
$x=$0.105. The strong renormalization of the SG as a function of
underdoping appears to be a generic feature of HTSC and might
signal a profound modification of the underlying electronic
excitations as the system approaches its Mott insulating phase.
In particular, we are not dealing with a simple damping or smearing
of the gap. The cusp upon heating through $T^{\ }_{c}$ implies that the
gap remains well-defined as it closes in energy, which also
indicates that the SG is directly coupled to the SC-order
parameter.

This work was supported by the Swiss National Science Foundation
(through NCCR, MaNEP, and grant Nr 200020-105151) and the Ministry
of Education and Science of Japan. Work in London was supported by a
Wolfson Royal Society Research Merit Award.


\begin{thebibliography}{99}
\bibitem{footnote}
The imaginary part of the magnetic susceptibility of a weakly
interacting gas of SC quasiparticles is calculated within the
random phase approximation. The fermiology parameters are chosen
to reproduce the $T$ dependence of the INS response at optimal
doping in LSCO (see Ref. \cite{softening d-wave slope}), with the
Ansatz
$\Delta(\mathbf{Q},T)=\Delta^{\ }_{0}({\mathbf{Q}})\sqrt{1-(T/T^{\ }_{c})^4}$
and $\Delta^{\ }_{0}(0,\pi)=10\ \mathrm{meV}$ for the $T$ dependence of
the $d$-wave SC gap.
\end{thebibliography}
\end{document}